\documentclass{iau}
\usepackage{graphicx}

\def\teamvipers{U. Abbas, C. Adami, S. Arnouts, J. Bel, M. Bolzonella, D. Bottini, E. Branchini, A. Burden, A. Cappi, J. Coupon, O. Cucciati, I. Davidzon, G. De Lucia, S. de la Torre, C. Di Porto, P. Franzetti,  A. Fritz, M. Fumana, B. Garilli, L. Guzzo, P. Hudelot, O. Ilbert, A. Iovino, J. Krywult, V. Le Brun, O. Le F{\`e}vre,  D. Maccagni, K. Ma{\l}ek, A. Marchetti, C. Marinoni, F. Marulli, H. J. McCracken, Y. Mellier, L. Moscardini, R. C. Nichol, L. Paioro, J. A. Peacock, W. J. Percival, M. Polletta, A. Pollo, M. Scodeggio, L. A. M. Tasca, R. Tojeiro, D. Vergani, G. Zamorani and A. Zanichelli}

\title[SCCC 21~~A high-dimensional look at VIPERS galaxies] 
{A high-dimensional look at VIPERS galaxies}

\author[Benjamin R. Granett \& the VIPERS Team]   
{Benjamin R. Granett$^1$
 \and the VIPERS Team$^2$}

\affiliation{$^1$INAF OA Brera, \\ Via E. Bianchi 46, 23807 Merate, Italy \\ email: {\tt ben.granett@brera.inaf.it} \\[\affilskip]
$^2$The VIPERS Team: \teamvipers \\Web address: {\tt http://vipers.inaf.it}}

\pubyear{2014}
\volume{306}  
\pagerange{0--0}
\jname{Statistical Challenges in 21st Century Cosmology}
\editors{A.C. Editor, B.D. Editor \& C.E. Editor, eds.}
\begin{document}

\maketitle

\begin{abstract}
We investigate how galaxies in VIPERS (the VIMOS Public Extragalactic Redshift Survey) inhabit the cosmological density field by examining the correlations across the observable parameter space of galaxy properties and clustering strength. The high-dimensional analysis is made manageable by the use of group-finding and regression tools. We find that the major trends in galaxy properties can be explained by a single parameter related to stellar mass. After subtracting this trend, residual correlations remain between galaxy properties and the local environment pointing to complex formation dependencies. As a specific application of this work we build subsamples of galaxies with specific clustering properties for use in cosmological tests.
\end{abstract}

\firstsection 
\section{Parameter space of VIPERS galaxies}
The VIMOS Public Extragalactic Redshift Survey (VIPERS) (\cite[Guzzo \etal\, 2014]{Guzzo}; \cite[Garilli \etal\, 2014]{Garilli}) is targeting 100k galaxies with spectroscopy at redshifts z=0.5 to 1.2. With complementary photometric coverage from UV to IR and morphological measures from CFHTLS, this unique dataset is providing unprecedented detail on the properties of galaxies and the formation of structure at intermediate redshifts.

We examine the correlations across the parameter space of observable and derived properties with special focus on spatial clustering.  We adopt a volume limited sample at $0.5 < {\rm z} < 1$ with luminosity cut $M_B < -19.5 - {\rm z}$ giving 15 494 galaxies.  We consider the following set of 12 parameters:
(1) \underline{$M_{\star}$}:   stellar mass;
(2)  \underline{$U-V$}: rest-frame spectral slope in blue;
(3) \underline{$M_V$}:	rest-frame V magnitude;
(4) \underline{$\phi$}:	PCA 1 spectral index;
(5) \underline{$D4000$}:	Balmer break depth;
(6) \underline{$OII EW$}:	OII line equivalent width;
(7) \underline{$OII Flux$}:	OII line apparent flux;
(8) \underline{$R-K$}:	rest-frame spectral slope in red;
(9) \underline{$\theta$}:	PCA 2 spectral index;
(10) \underline{$sersic$}:	Sersic morphological index;
(11) \underline{$\gamma$}:	Clustering slope at $1 < r < 30 {\rm Mpc}/h$;
(12) \underline{$A$}:	Clustering amplitude at $5 < r < 30 {\rm Mpc}/h$.

The spectral principal component analysis is described in \cite{marchetti}; the derivations of the rest-frame galaxy properties and stellar mass are given in \cite{davidzon} and \cite{fritz}; spectral line measurements are described in \cite{garilli}; the spatial clustering properties are described in \cite{delatorre} and \cite{marulli}; the galaxy morphologies are measured in Krywult \etal (in preparation).  
These parameters represent diverse quantities with different units.  To measure separations between points in parameter space, we first apply a Gaussianisation transform to the distributions.

\begin{figure}[t]
\begin{center}
\includegraphics[width=4in]{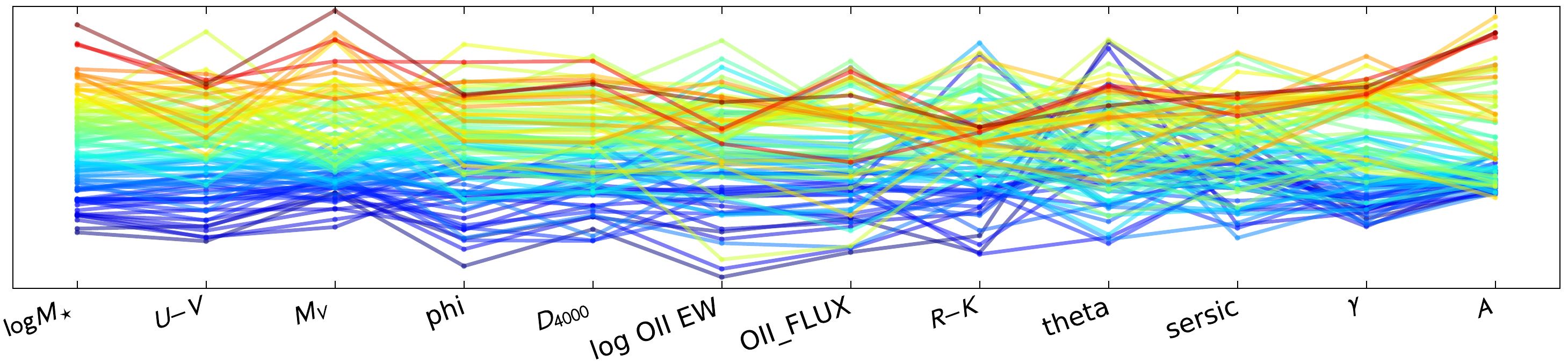} 
 \includegraphics[width=3in]{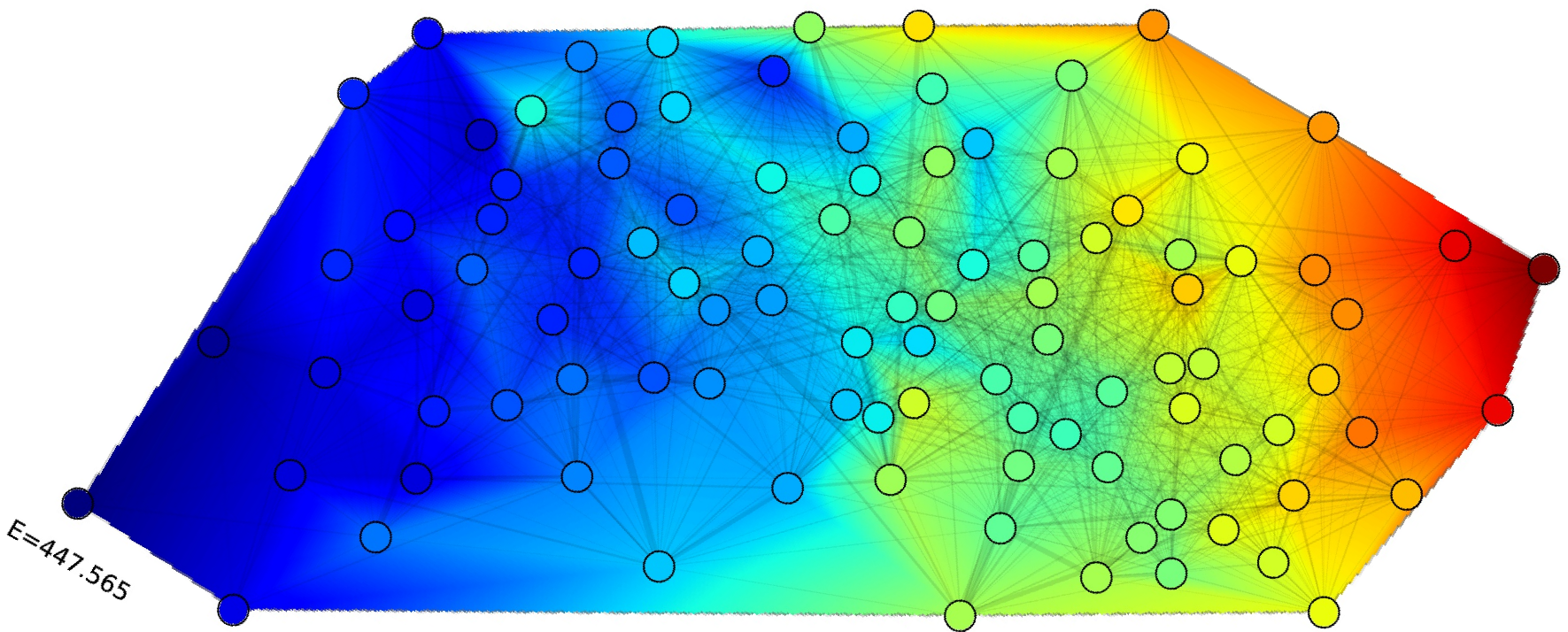} \includegraphics[width=1.5in]{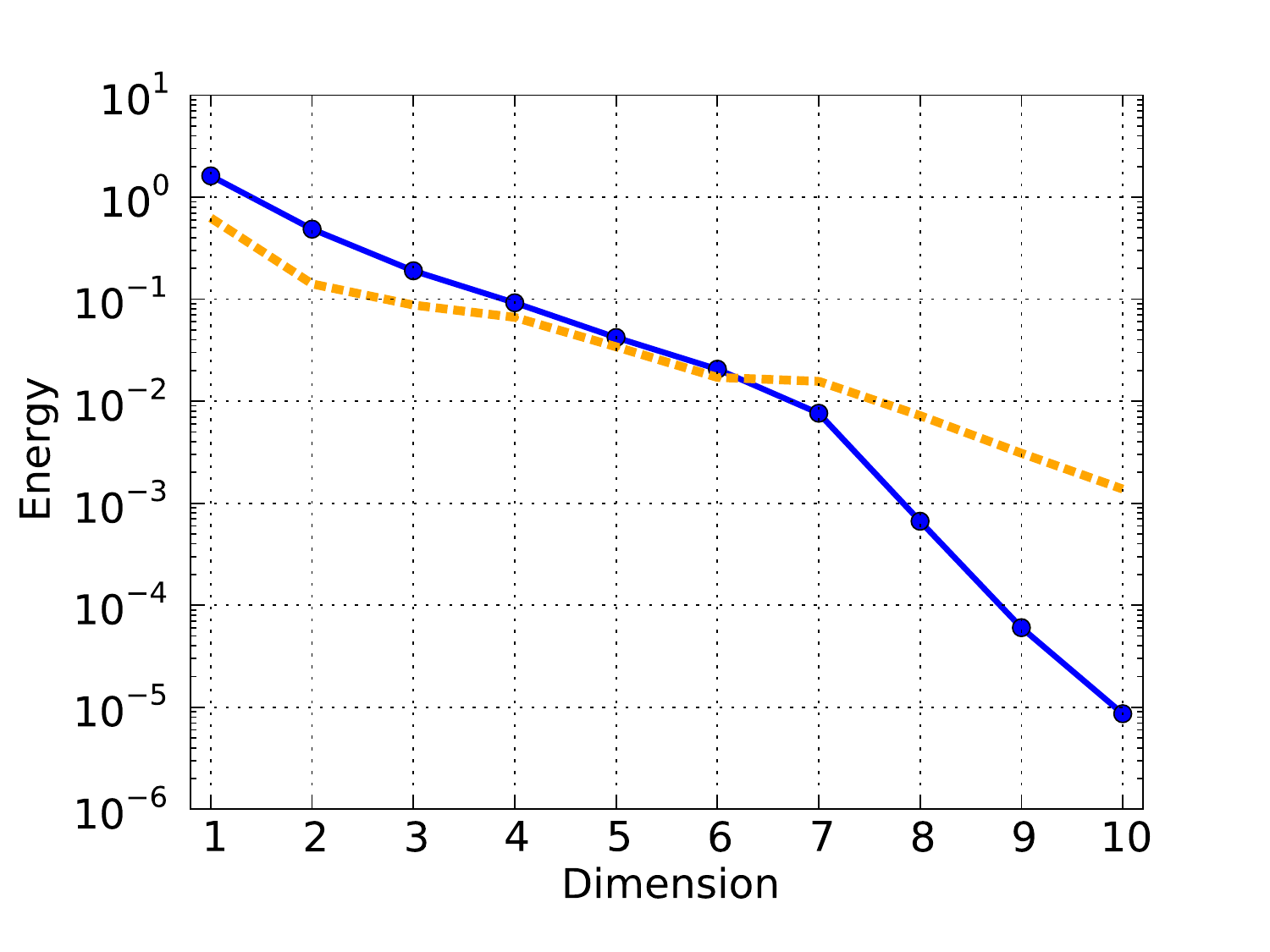} 

 \caption{{\it Top:} A parallel-coordinates plot of group centres in the 12D space coloured according to $M_\star$. The order of colours is mostly preserved across the plot indicating strong correlations.
 {\it Left:}
 A graph of the connections between galaxies coloured by $M_\star$.  Neighbouring group centres are connected by an edge with length given by their Gaussianised separation.  The graph is flattened onto the page allowing edges to stretch and compress while similar groups remain near to each other.  The uniform colour gradient shows that sorting by a single parameter, $M_\star$, can organise the groups into a one-dimensional sequence. 
 {\it Right:}
 The solid line shows the graph energy with dimension D.   The energy follows a power-law trend; the knee at $D=7$ points to the underlying dimension. For comparison, the eigenvalues are overplotted (orange dashed line).
 }
   \label{fig:dim}
\end{center}
\end{figure}

We group the galaxies by properties to further reduce the dimensionality of the dataset.  We apply the k-means group-finding algorithm to determine 100 group centres.  This algorithm partitions the space into subgroups with approximately equal size.  The projected correlation function of each subgroup is measured to derive the $\gamma$ and $A$ parameters.  The $\gamma$ and $A$ values are then assigned to individual galaxies by extrapolating between the group centres. 
What is the underlying dimensionality of the dataset?  We show in Fig. \ref{fig:dim} that a single parameter such as stellar mass can explain
the primary trends.  

\section{Correlations in parameter space}

\begin{figure}[t]
\begin{center}
\includegraphics[width=4.5in]{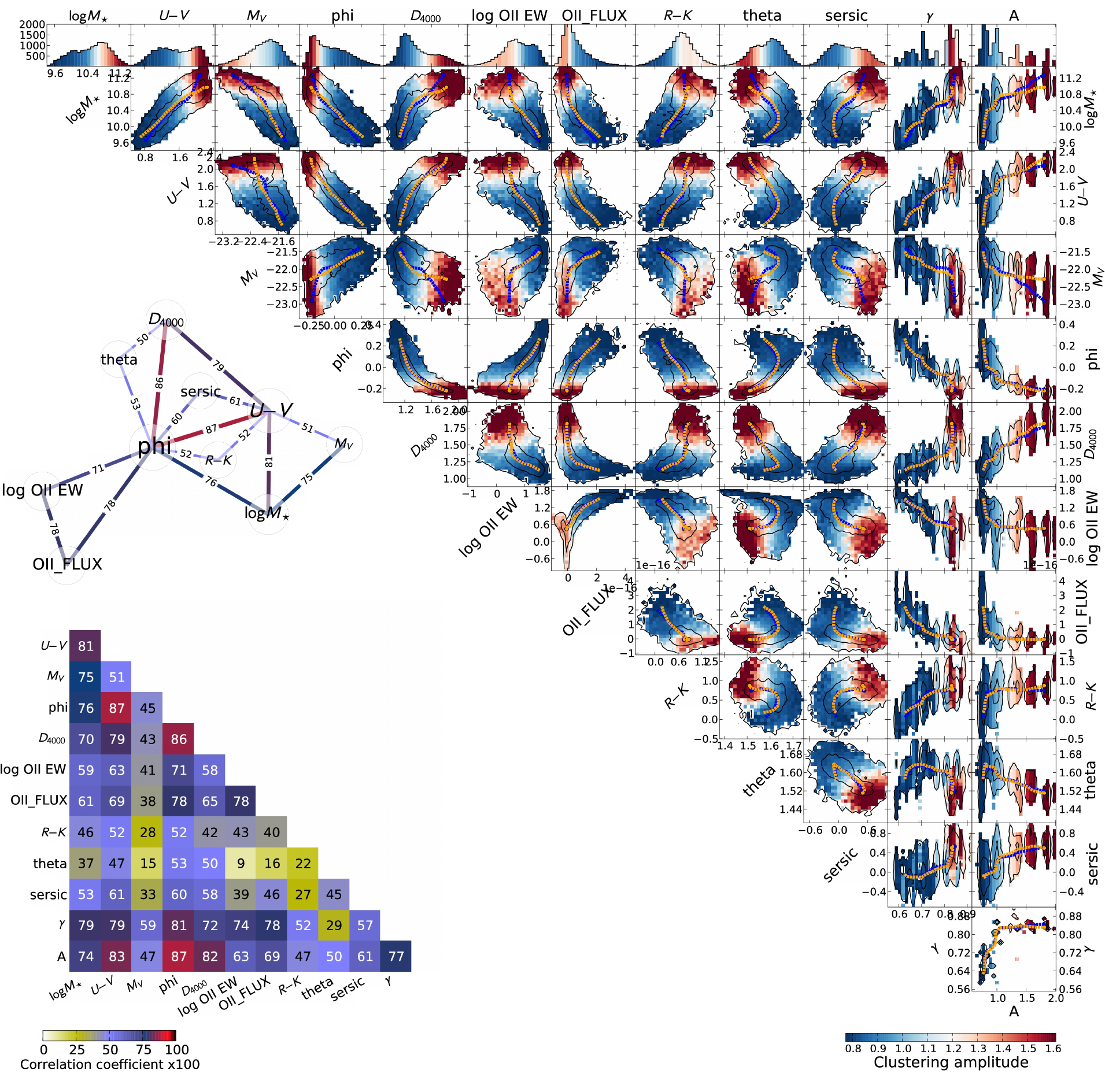} 
 \caption{Connections between parameters quantified by Spearman's rank correlation coefficient. Stellar mass, luminosity and D4000 are connected but separated from line strength.  The PCA $\phi$ parameter emerges as one of the most informative parameters.
The panels show the joint distributions of parameters coloured by clustering amplitude A.  The contours contain 50\% and 90\% of galaxies.
Orange dashed line  gives the trend with stellar mass while blue dashed line gives the trend with $M_V$ showing that mass and luminosity can account for the primary trends.}
   \label{fig:big}
\end{center}
\end{figure}

We further examine the correlations between parameters.  Fig. \ref{fig:big} shows the joint  distributions between all pairs of parameters.
The bimodal nature of galaxies is reflected in most of the properties considered.  Most striking is the clustering amplitude $A$ that
characterises how galaxies are distributed over the density field.  Considering the two locii of galaxies, the reddest, luminous and massive group is also highly clustered signifying dense environments, while the bluer galaxies are more uniformly distributed and show weaker clustering slopes $\gamma$.

To investigate the factors that influence the galaxy properties we subtract the median trend with $M_\star$ from each parameter (not shown here).
Once removed, the correlations are greatly reduced.  In particular, correlations with $M_V$ are reduced since $M_V$ is degenerate with stellar mass.  However, we find that colour parameters $U-V$, $\phi$ and $D4000$ remain strongly correlated and still relate to clustering.

Stellar mass is a major driver of galaxy properties, but it cannot explain everything.  After subtracting trends with stellar mass, residual correlations exist reinforcing the conclusion that environment must play an important role.  
This is highlighted by the dependencies seen in the correlation function since the clustering properties may be related directly to how galaxies are distributed within underlying dark matter clumps.  We see that the dependencies on galaxy properties are complex as they encode the subtle mechanisms relating the formation of galaxies to the large-scale density field.

\small
We acknowledge the ESO staff for the management of service
observations. We are grateful to M. Hilker for his constant help and support.  
BRG acknowledges support of the European Research
Council through the Darklight ERC Advanced Research Grant (\#
291521).  A complete list of the funding agencies that support  VIPERS may be found
at http://vipers.inaf.it.

\end{document}